\documentclass[a4paper,11pt]{article}
\usepackage{jinstpub} 
\usepackage{lineno}

\newcommand{\etal}{\rm et~al.}

\title{SPIROS: Streamlined, Precise, Intuitive, and Rapid Optical Simulator for particle physics detectors}

\author{Tatsuya Kikawa}

\affiliation{Department of Physics, Kyoto University, Kyoto, Kyoto 606-8502, Japan}

\emailAdd{kikawa.tatsuya.6e@kyoto-u.ac.jp}

\abstract{
This paper presents SPIROS (Streamlined, Precise, Intuitive, and Rapid Optical Simulator), a dedicated optical simulation tool developed for the design and analysis of particle physics detectors. Unlike general-purpose frameworks such as GEANT4, SPIROS offers a lightweight simulation engine and a user-friendly interface optimized for optical processes, including scintillation, Cherenkov emission, and photon transport with reflection, refraction, scattering, absorption, and detection. Detector geometries can be directly imported from 3D CAD models, and all configurations including materials, surfaces, sources, and sensors are specified via a single human-readable input file. Validation against GEANT4 shows excellent agreement in photon generation and propagation behaviors, while benchmark tests demonstrate that SPIROS runs more than two times faster for typical detector configurations. The software has already been applied to multiple neutrino experiments, including T2K, NINJA, and AXEL, for detector design, performance studies, and optimization. SPIROS is open-source and freely available at https://github.com/tkikawa/spiros.
}


\begin{document}
\maketitle
\flushbottom

\section{Introduction}\label{sec_intro}

Optical simulations play a crucial role in the design and optimization of detectors used in particle physics experiments, particularly those involving scintillation or Cherenkov light detection. Accurate modeling of photon propagation, reflection, scattering, and detection is essential for understanding detector response and improving reconstruction performance.

GEANT4~\cite{bib_g4_1,bib_g4_2,bib_g4_3}, a widely used toolkit for simulating the passage of particles through matter, includes capabilities for detailed optical photon tracking. However, configuring GEANT4 for optical simulations is often highly complex and time-consuming. It requires in-depth knowledge of the framework and substantial effort to define the detector geometry, materials, surfaces, and physics processes involved. As a result, performing iterative design studies or rapid prototyping with GEANT4 can be prohibitively inefficient in practice.

On the other hand, several commercial optical simulation tools exist, but they are typically tailored to applications in engineering or medical imaging, and lack essential features or flexibility required for particle physics experiments. Moreover, their closed-source nature and limited support for custom detector geometries and physics models pose significant challenges.

To address these issues, I have developed SPIROS, a streamlined, precise, intuitive, and rapid optical simulator specifically designed for particle physics experiments. This tool enables rapid simulation of optical processes through a lightweight and highly optimized execution engine. It offers an intuitive user experience by supporting human-readable configuration files, minimal setup requirements, and direct import of detector geometries from 3D CAD models. The framework’s streamlined architecture facilitates fast iteration and prototyping, while its physically accurate modeling of optical interactions ensures precise and reliable results suitable for both design and analysis tasks. Simulation results can be explored using a built-in interactive event display for visual inspection of the optical behavior, or exported to data files for detailed quantitative analysis.
SPIROS is released as open-source software and is freely available at https://github.com/tkikawa/spiros, along with a user manual and several example configuration files.

This paper presents the design philosophy, software architecture, and computational performance of SPIROS. Its simulation results are validated through comparison with GEANT4, and its applicability is demonstrated through use in the design and optimization of actual particle physics experiments.
The remainder of this paper is organized as follows.
Section~\ref{sec_implementation} describes the design and implementation of SPIROS, including its architecture, configuration system, and geometry handling.
Section~\ref{sec_validation} presents validation results through a comparison with GEANT4 simulations.
Section~\ref{sec_performance} evaluates the computational performance of SPIROS relative to GEANT4.
Section~\ref{sec_application} demonstrates applications of SPIROS in actual particle physics experiments.
Finally, conclusions are given in Section~\ref{sec_conclusion}.

\section{Design and Implementation}\label{sec_implementation}

\subsection{Architecture Overview}
The simulation engine is implemented in C++ and leverages two key external libraries: ASSIMP (Open Asset Import Library)~\cite{bib:assimp} and ROOT (a data analysis framework developed by CERN)~\cite{bib:root_nim, bib:root}. ASSIMP is used to import geometry from various 3D CAD file formats, allowing users to define detector layouts using familiar design tools such as SolidWorks or Autodesk Inventor. ROOT is employed for both simulation output and event display, providing structured data storage and interactive visualization capabilities, and is also used internally to generate Landau-distributed fluctuations for scintillation modeling.

The software supports both optical-photon and charged-particle source modes. Photon transport includes refraction, reflection (specular and diffuse), absorption (bulk and surface), scattering, and detection, while charged particles can produce scintillation or Cherenkov light based on material properties. These processes are handled through a modular transport engine that governs photon and particle behavior within the geometry.

The Mersenne Twister pseudorandom number generator~\cite{bib_mt} is used to sample physical processes efficiently, ensuring statistically reliable and reproducible simulations.

\subsection{Implementation of Optical and Particle Physics Processes}

\subsubsection{Optical Photon Transport}\label{subsec_photon_trans}
Each optical photon is propagated step-by-step through the detector geometry based on geometric boundary tracking and interaction probabilities.
The polarization state of each photon is also tracked throughout the simulation and updated appropriately at each optical interaction.
At material boundaries, the behavior of the photon is determined by the surface properties of the material:

\begin{itemize}
  \item Normal medium surfaces are treated using Fresnel equations to compute the reflection and refraction probabilities based on the angle of incidence and refractive indices of adjacent materials, with the direction of refraction determined by Snell’s law. These interactions include polarization-dependent reflection and transmission coefficients.
  \item Mirror surfaces reflect photons specularly.
  \item Diffuser surfaces apply Lambertian (cosine-distributed) reflection~\cite{bib_lambert}.
  \item Absorber surfaces terminate photons upon contact.
  \item Detector surfaces also terminate photons, but may register them as "detected" for output analysis based on user-defined detection efficiency.
  \item Converter surfaces randomize the direction of photons isotropically upon entry into the material, effectively simulating the absorption and re-emission processes as in wavelength shifting.
\end{itemize}

SPIROS also supports material mixtures, where the boundary behavior is defined by a probabilistic combination of two or more material types. Upon boundary contact, the photon undergoes a Monte Carlo selection based on these weights and then follows the corresponding interaction model. For example, a surface may be defined as 70\% mirror and 30\% absorber, meaning that photons incident on the surface will undergo specular reflection with 70\% probability and absorption with 30\% probability.
This mechanism enables simulation of partially coated surfaces, imperfect reflectors, half-mirrors, or rough surfaces without explicit geometric modeling of microstructure.

Within each material, bulk optical properties are modeled probabilistically.
SPIROS assumes wavelength-independent optical parameters for simplicity, which is sufficient for many detector design and optimization studies.
The step length to the next interaction is sampled based on exponential attenuation and scattering lengths. If the step results in absorption or scattering, the photon's trajectory is updated accordingly. Rayleigh scattering is enabled by default when a non-zero scattering length is provided, and Mie scattering can optionally be activated with specifying the asymmetry factor.

This transport continues until the photon either exits the simulation world, is absorbed, is detected, or exceeds the user-defined limit on the number of reflections.

\subsubsection{Charged Particle Handling and Secondary Photon Generation}
When operating in the charged-particle source mode, SPIROS simulates the passage of primary charged particles through the geometry. Their path is used to estimate the expected number of secondary optical photons produced by scintillation and Cherenkov radiation.

For scintillation, the expectation is calculated using the energy loss per unit length (d$E$/d$x$) from the Bethe–Bloch or Berger-Seltzer equation with shell and density corrections, the particle's path length, and a user-defined scintillation yield specified as the number of photons produced per unit energy deposited. The actual number of photons is then determined by applying a Landau distribution to model fluctuations in energy deposition, followed by Poisson sampling.
In practice, properties of typical scintillators such as NaI(Tl) and CsI(Tl) are predefined based on Ref.~\cite{bib:knoll}.

In the case of Cherenkov radiation, the expected number of photons is computed from the Frank–Tamm formula~\cite{bib:frank_tamm} based on the particle's speed and the refractive index of the medium, and the actual count is drawn from a Poisson distribution.
The emission positions are sampled along the particle trajectory, with isotropic angular distributions for scintillation and cone-like distributions for Cherenkov photons, where the emission angle is determined by the particle’s speed and the refractive index.
Cherenkov photons are initialized with 100\% linear polarization perpendicular to the emission plane, consistent with physical theory.
Each secondary photon is then transported through the geometry using the optical photon transport algorithm described in Section~\ref{subsec_photon_trans}, including interactions such as boundary reflection, refraction, absorption, and scattering.

\subsubsection{Time Handling}
SPIROS tracks the propagation time of all particles for timing-based analyses. The generation time of the primary particle is set to zero, and the photon arrival time is calculated by integrating the path length over the local light speed determined by the refractive index.
In the charged-particle source mode, the photon time includes the charged particle’s flight time to the emission point, a delayed scintillation emission time sampled from an exponential distribution, and the photon’s travel time through the geometry.

\subsection{Configuration and Geometry Handling}
\begin{figure}[htbp]
    \begin{center}
        \includegraphics[width=0.9\linewidth]{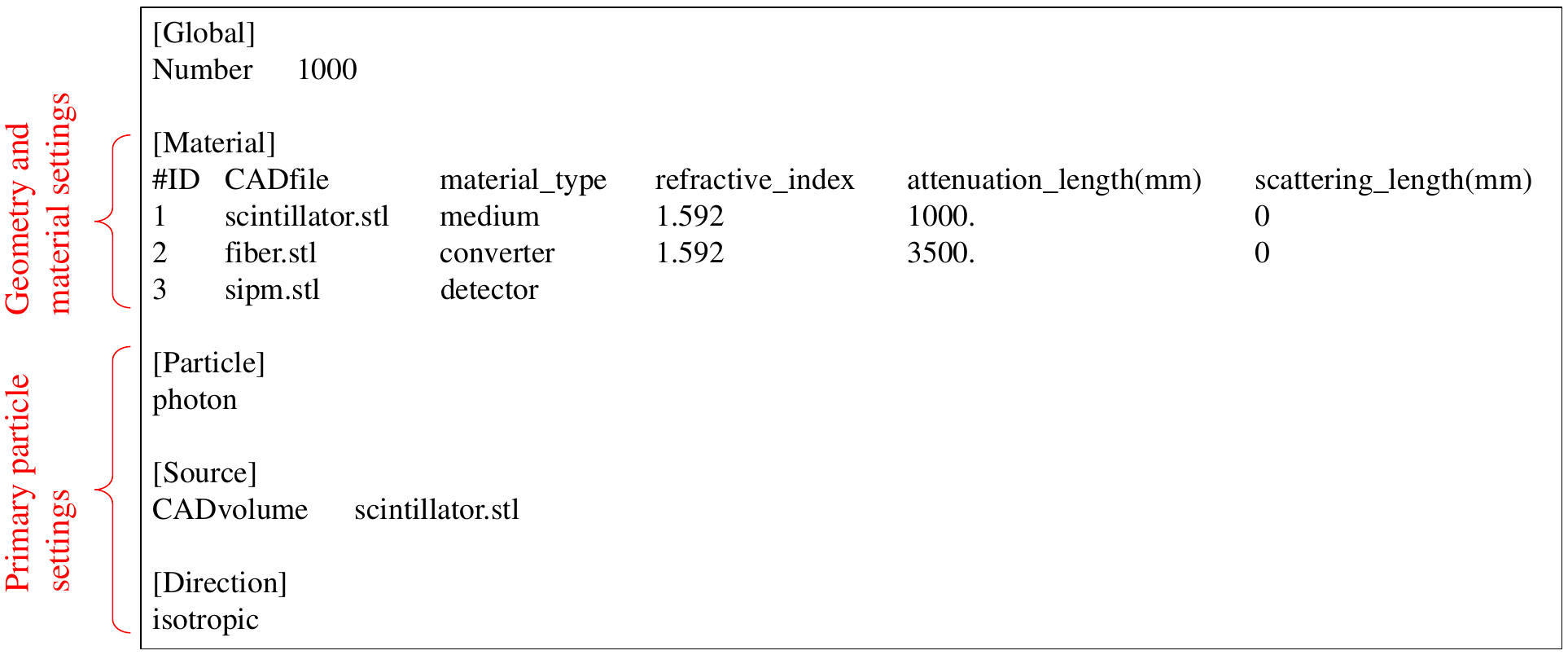}
         \caption{
            Example of an input configuration file. 
         }
         \label{fig:input}
    \end{center}
\end{figure}

SPIROS is designed for usability and rapid prototyping. All simulation parameters ranging from global settings to detailed material definitions and particle source configurations are specified in a single human-readable input file.
An example input file used to simulate the generation of 1,000 photons uniformly and isotropically within a scintillator, followed by their transport through a wavelength-shifting (WLS) fiber and detection by silicon photomultipliers (SiPMs), is shown in Figure~\ref{fig:input}.
This approach allows users to configure and execute simulations without the need for recompilation or code modification, facilitating iterative detector studies and parameter scans.

\subsubsection{Geometry Import}\label{subsec_geometry}
Detector geometry is defined via external 3D CAD files, which are imported using the ASSIMP library. A commonly used format is STL, which represents the detector's surfaces as triangle meshes and is widely supported by CAD software.
This approach allows users to incorporate detector models directly from CAD design environments, eliminating the need to manually recreate the geometry and streamlining the simulation workflow.
Each distinct material region in the detector is associated with a CAD file and assigned a material type such as normal medium, mirror, or diffuser, as described in Section~\ref{subsec_photon_trans}.

To accelerate photon tracking in mesh-based geometries, SPIROS organizes the imported triangle meshes using a bounding volume hierarchy (BVH)~\cite{bib:bvh}. In this structure, triangles are grouped hierarchically into nodes, each associated with an axis-aligned bounding box (AABB) that encloses the triangles in the node. During photon propagation, intersection tests are first performed against these bounding boxes. If the photon trajectory does not intersect a given bounding box, all triangles contained in that node are skipped without performing individual triangle intersection tests. This hierarchical culling greatly reduces the number of triangle intersection calculations required for complex geometries.

\subsubsection{Material and Particle Configuration}
The input file allows users to flexibly define the optical properties of each material, including the material type, refractive index, and attenuation and scattering lengths. Materials can also exhibit complex surface behaviors such as partial reflectivity or diffuse scattering, and may be defined as probabilistic mixtures of multiple types, as described in Section~\ref{subsec_photon_trans}

Primary particle generation is similarly configurable, allowing users to define source information including particle type, energy, position, direction, and polarization using simple keywords.
When operating in the optical-photon source mode, the degree of polarization can be specified for each emitted photon. In the charged-particle source mode, users can configure the particle type and initial energy.
Both CAD-based and analytical source geometries (e.g., boxes, spheres, cylinders) are supported.
A variety of angular distributions are available, including isotropic emission, directional beams, Lambertian profiles commonly used for LED modeling~\cite{bib_lambert}, and distributions proportional to cos$^2\theta$, relevant for cosmic-ray simulations.
For both source position and angular distributions, a custom mode is also available, allowing users to define arbitrary spatial or angular functions tailored to specific use cases.

\subsubsection{Output and Visualization}
The results of a SPIROS simulation are stored in ROOT format using the TTree structure, containing detailed information on photon trajectories, interaction history, timing, and termination conditions.
In the charged-particle source mode, information on the primary charged particles is also recorded and associated with the corresponding secondary photons.
These outputs support both quantitative analysis such as evaluation of photon detection efficiency or time profiles and visual inspection of event-level behavior.

\begin{figure}[htbp]
    \begin{center}
        \includegraphics[width=0.45\linewidth]{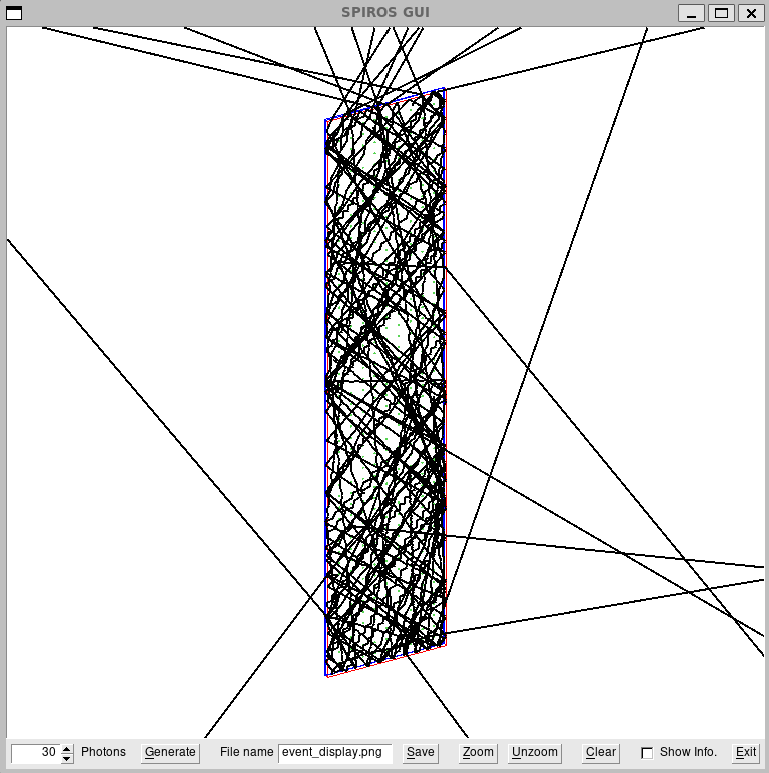}
         \caption{
            Built-in interactive event display.
         }
         \label{fig:evtdisp}
    \end{center}
\end{figure}

SPIROS also provides an optional graphical display mode for real-time event visualization as illustrated in Figure~\ref{fig:evtdisp}. It allows users to inspect the detector geometry, specify the number of primary particles, initiate the simulation interactively, and observe the resulting tracks of optical photons and charged particles overlaid on the geometry. The interface supports intuitive interactions such as rotation and zooming, and rendered views can be exported as image files.

\section{Validation Against GEANT4}\label{sec_validation}

To validate the physical accuracy of SPIROS, simulation results were compared against those obtained from GEANT4 (v11.3.2) across several representative configurations. All simulations were conducted using consistent geometry, material, and optical parameters in both frameworks.
In GEANT4, the geometry was constructed using native solid classes such as G4Box and G4Tubs, whereas in SPIROS, the geometry was defined using STL files exported from Autodesk Inventor with default mesh resolution. The impact of this modeling difference on simulation performance and accuracy is discussed in Section~\ref{sec_performance}.

\subsection{Scintillator and Photomultiplier Tube}\label{subsec_validation_scinti}
The first validation case involves a 50~mm $\times$ 50~mm $\times$ 500~mm plastic scintillator bar with a 40~mm diameter photomultiplier tube (PMT) coupled to one end. A 1~GeV muon beam is directed perpendicularly onto the 50~mm $\times$ 500~mm face of the scintillator. Default parameters include a light yield of 3,000 photons/MeV, attenuation length of 500~mm, Rayleigh scattering length of 300~mm, and PMT quantum efficiency of 20\%. Cherenkov radiation is disabled in this configuration.

\begin{figure}[htbp]
  \centering
  \begin{minipage}{0.47\textwidth}
    \centering
    \includegraphics[width=1.08\linewidth]{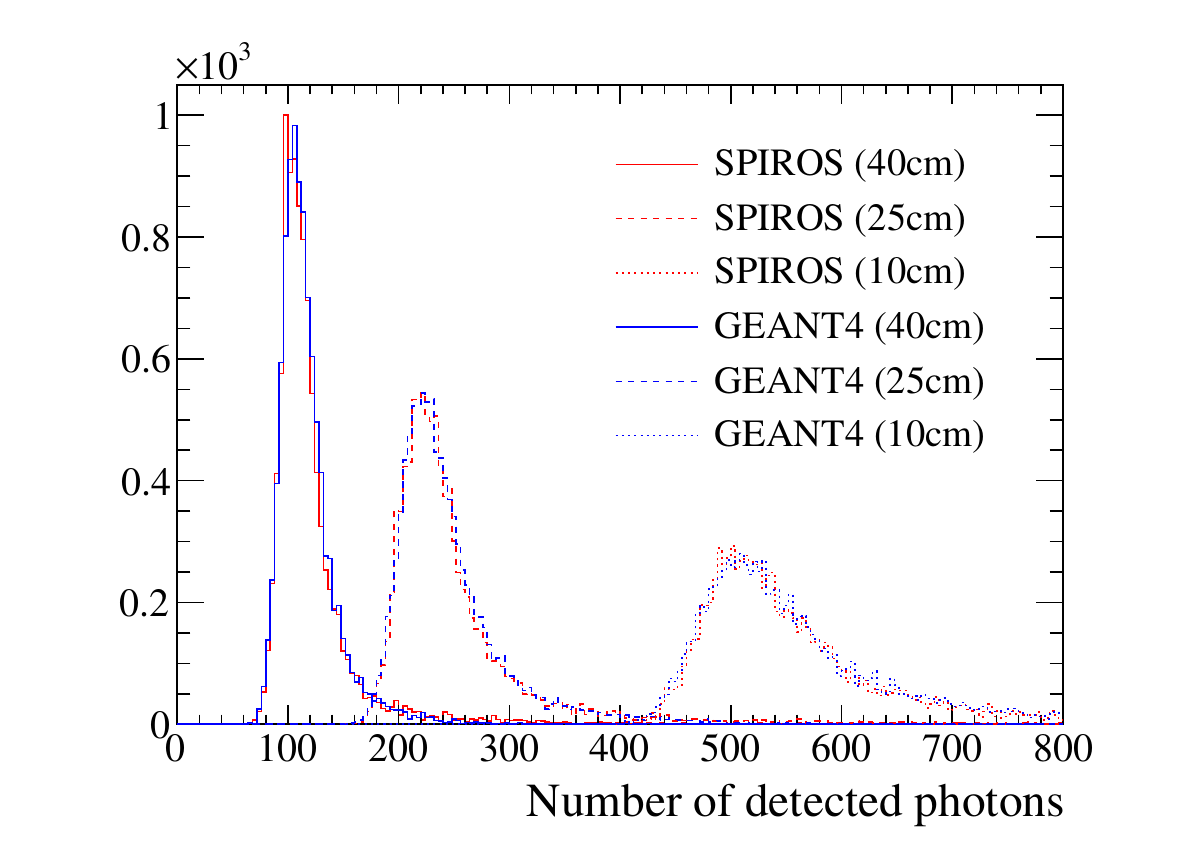}
    \caption{Distribution of detected photon counts for 1~GeV muons incident at 10, 25, and 40~cm from the PMT in a scintillator.}
    \label{fig:ly}
  \end{minipage}
  \hfill
  \begin{minipage}{0.47\textwidth}
    \centering
    \includegraphics[width=1.08\linewidth]{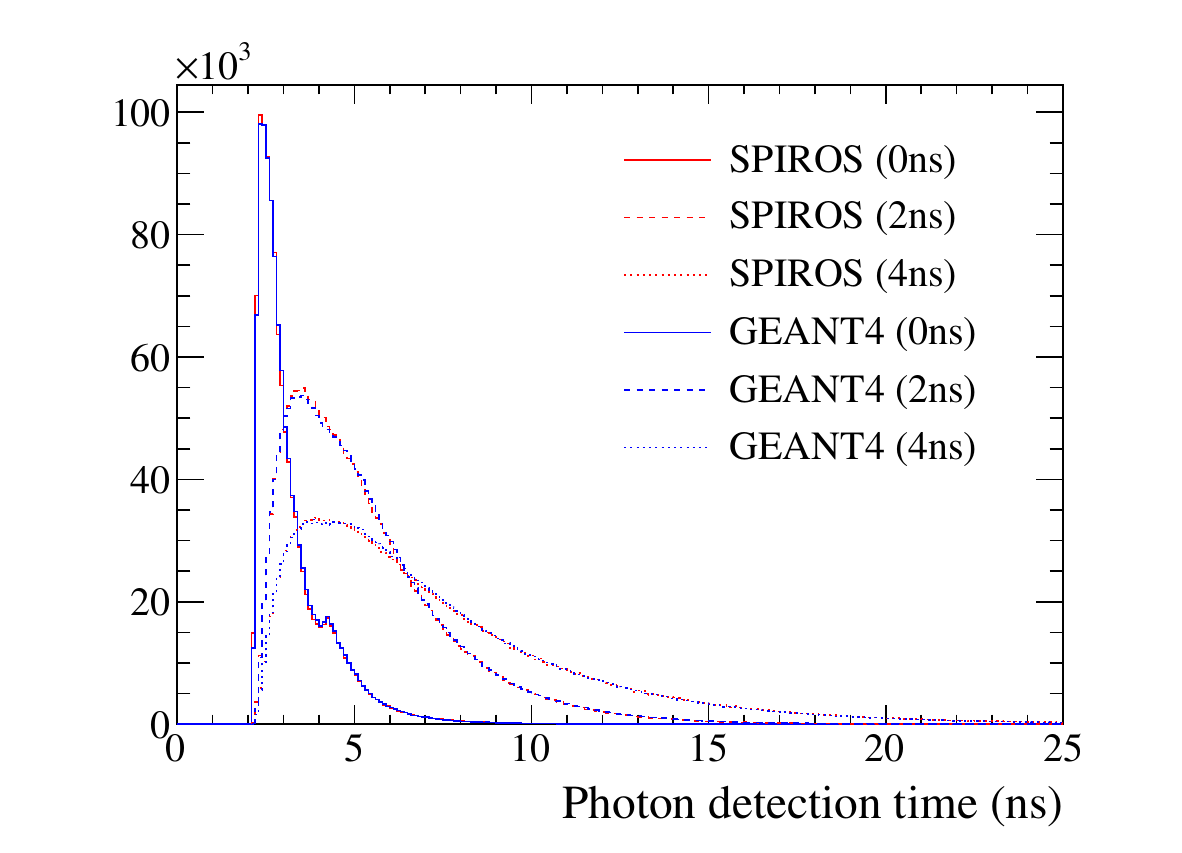}
    \caption{Photon detection time distributions for different scintillation decay times (0, 2 and 4~ns).}
    \label{fig:timing}
  \end{minipage}
\end{figure}

\begin{figure}[htbp]
  \centering
  \begin{minipage}{0.47\textwidth}
    \centering
    \includegraphics[width=1.08\linewidth]{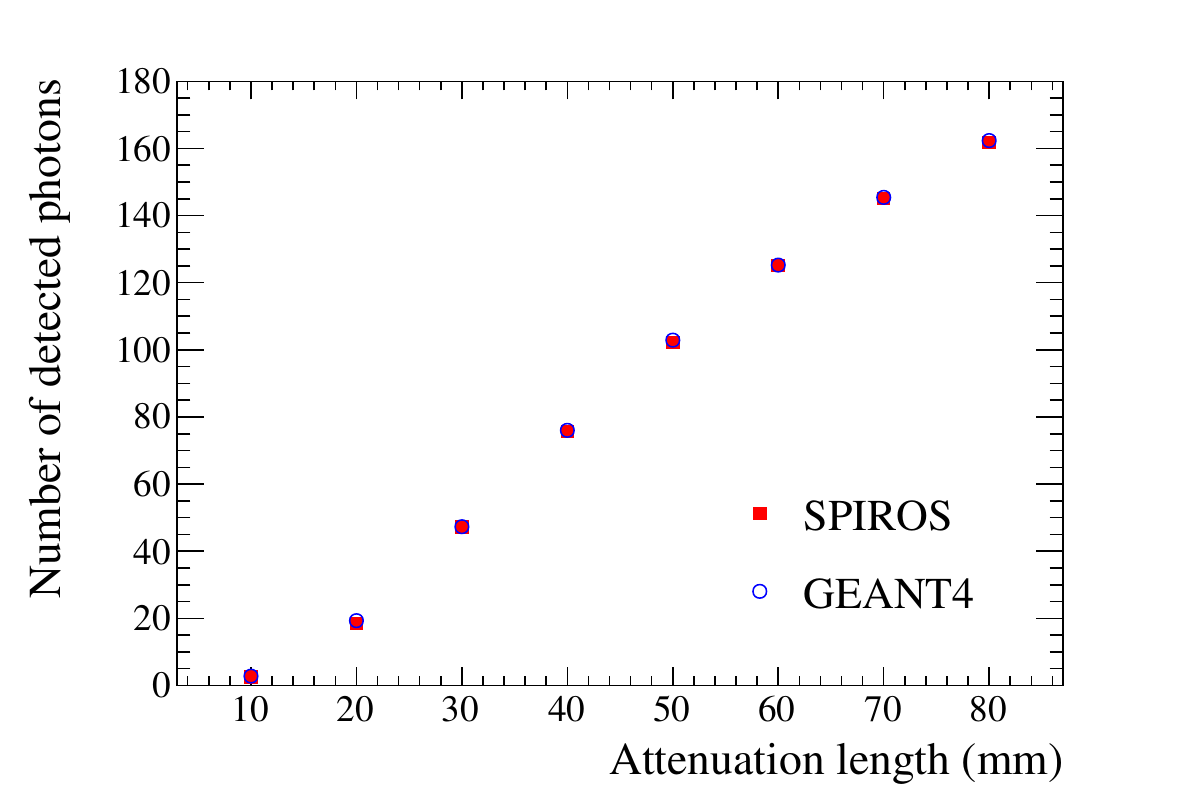}
    \caption{Dependence of the average number of detected photons on the attenuation length of the scintillator.}
    \label{fig:att}
  \end{minipage}
  \hfill
  \begin{minipage}{0.47\textwidth}
    \centering
    \includegraphics[width=1.08\linewidth]{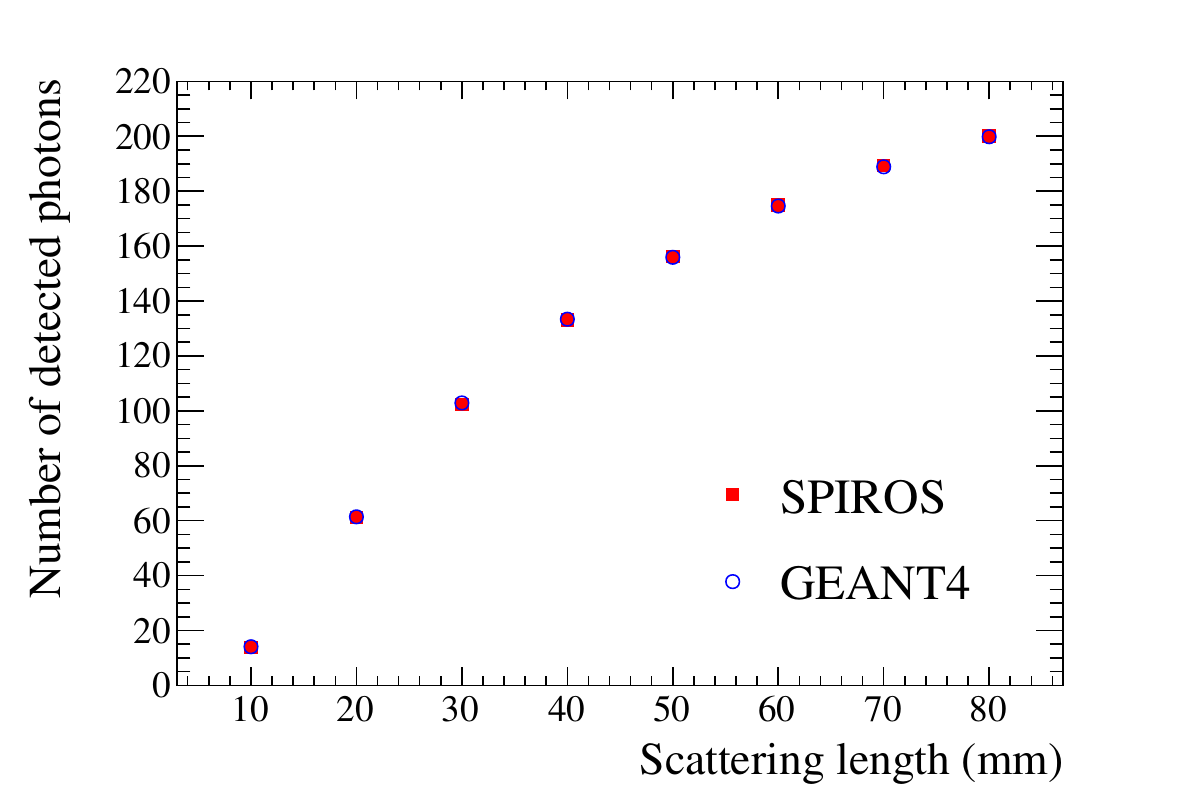}
    \caption{Dependence of the average number of detected photons on the Rayleigh scattering length of the scintillator.}
    \label{fig:scat}
  \end{minipage}
\end{figure}

Figure~\ref{fig:ly} compares the distributions of detected photon counts per muon event, evaluated over 10,000 events, for injection positions at 10, 25 and 40~cm from the PMT. Figure~\ref{fig:timing} shows the timing distribution of detected photons for scintillation decay times of 0, 2, and 4 ns. In addition, Figures~\ref{fig:att} and \ref{fig:scat} illustrate how the average detected light yield varies with different attenuation and scattering lengths, respectively.

The results show excellent agreement between SPIROS and GEANT4, confirming the correct implementation of scintillation photon generation, transport, absorption, scattering, and timing.

\subsection{Cherenkov Ring Imaging with Aerogel}\label{subsec_validation_cherenkov}
The second validation case focuses on Cherenkov radiation. A slab of aerogel with thickness 2~cm and refractive index 1.05 is illuminated by perpendicularly incident 4~GeV/c pions or kaons. A position-sensitive photon detector is placed 20~cm downstream to capture the Cherenkov ring image.
The photon detector is sensitive to wavelengths in the 300--500 nm range with quantum efficiency of 20\% by default.

\begin{figure}[htbp]
  \centering
  \begin{minipage}{0.47\textwidth}
    \centering
    \includegraphics[width=1.08\linewidth]{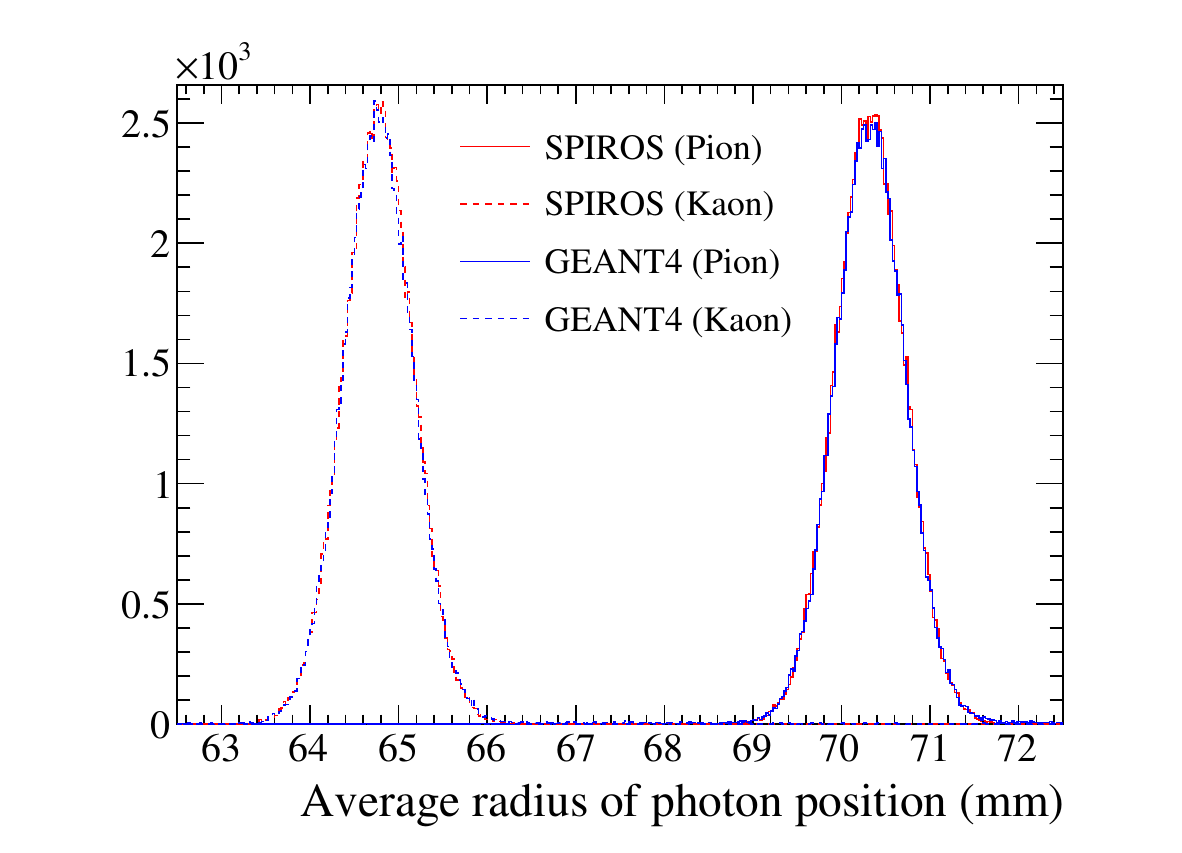}
    \caption{Average radius distribution of detected photon hits on a detector placed downstream of an aerogel radiator, for 4~GeV/c pions and kaons.}
    \label{fig:pid}
  \end{minipage}
  \hfill
  \begin{minipage}{0.47\textwidth}
    \centering
    \includegraphics[width=1.08\linewidth]{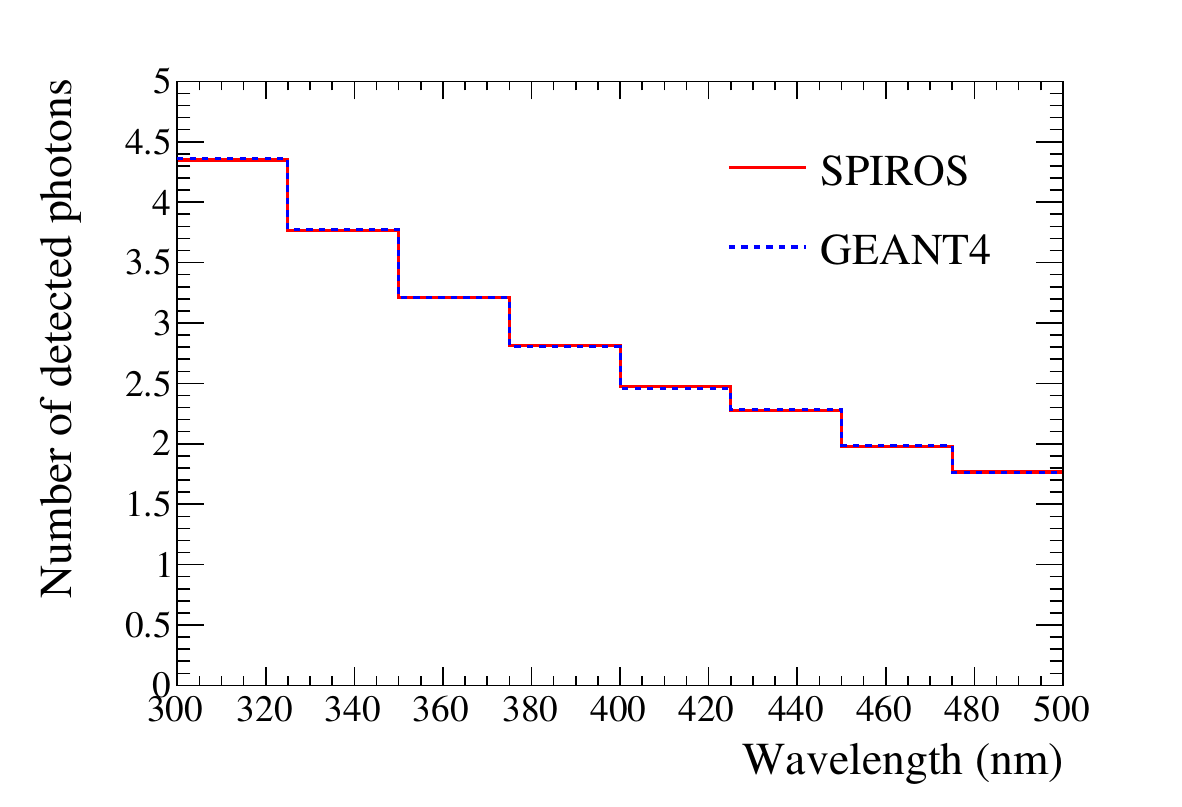}
    \caption{Wavelength spectra of detected Cherenkov photons for different detector sensitivity windows under 4~GeV/c pion irradiation.}
    \label{fig:wavelength}
  \end{minipage}
\end{figure}

Figure~\ref{fig:pid} shows the distribution of average radial distances of detected photon hits from the particle axis, calculated over 100,000 incident particles. Figure~\ref{fig:wavelength} represents the wavelength spectra of detected photons for various sensitivity windows under 4~GeV/c pion irradiation.

These results demonstrate strong agreement between SPIROS and GEANT4, validating the treatment of Cherenkov emission, angular distribution, wavelength filtering, and polarization handling.

\subsection{Wavelength-Shifting Fiber Test with LED}\label{subsec_validation_fiber}
The third case tests the optical-photon source mode with a WLS fiber. A 1~mm diameter, 2~m long fiber (core $n=1.59$, cladding $n=1.49$) is equipped with a SiPM on one end. An LED illuminates the fiber from the side at a distance of 5~mm, with a total emission of 10,000 photons. A Lambertian angular profile is used to model the directional emission from the LED and the attenuation length in the fiber is set to 3~m.

\begin{figure}[htbp]
  \centering
  \begin{minipage}{0.47\textwidth}
    \centering
    \includegraphics[width=1.08\linewidth]{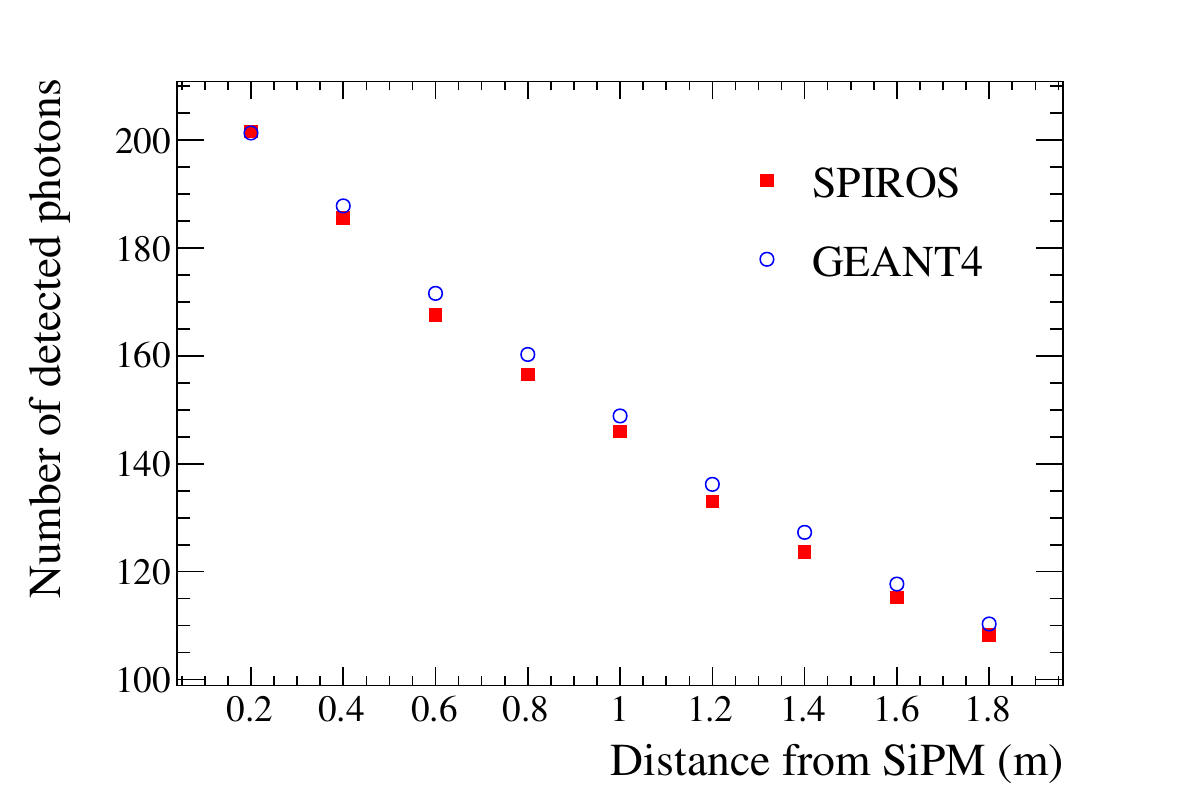}
    \caption{Average number of photons detected by the SiPM as a function of LED illumination position along a WLS fiber.\newline\newline}
    \label{fig:fiber_led}
  \end{minipage}
  \hfill
  \begin{minipage}{0.47\textwidth}
    \centering
    \includegraphics[width=1.08\linewidth]{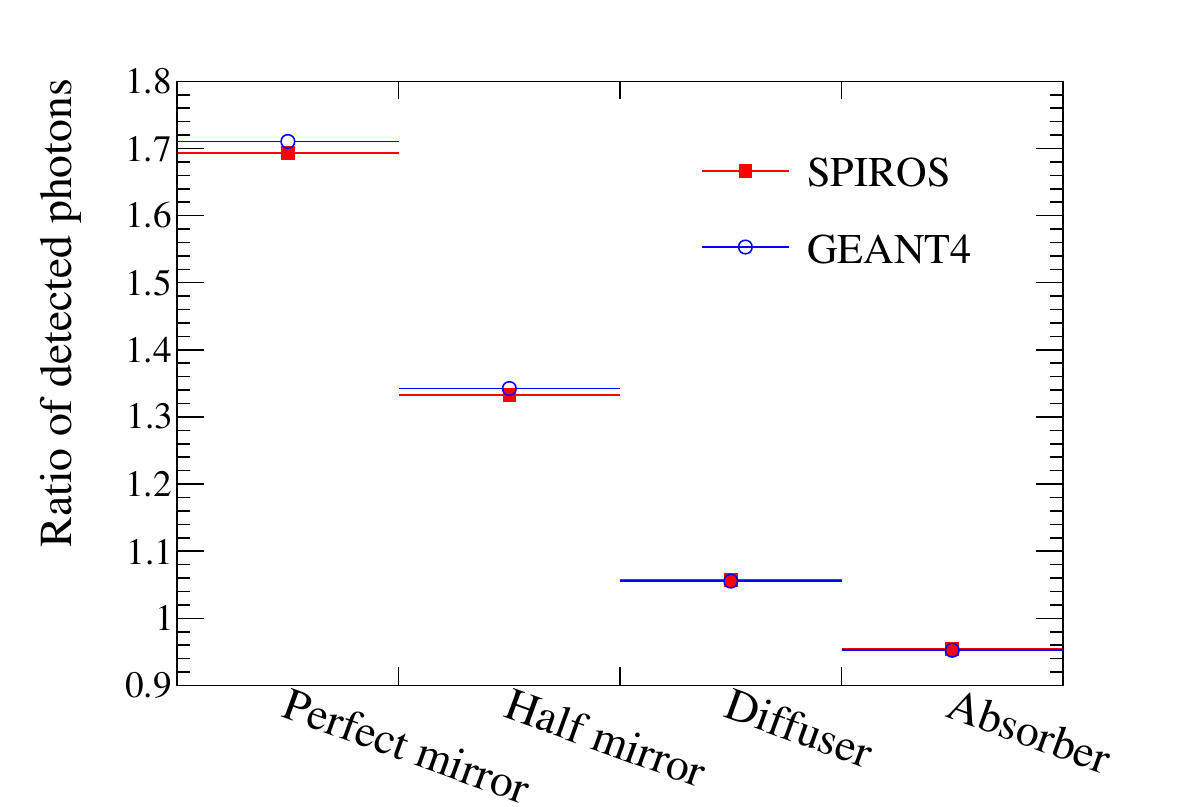}
    \caption{Relative number of photons detected by the SiPM with different surface treatments at the far end of the WLS fiber, normalized to the untreated case when the LED–SiPM distance is 1.6~m.}
    \label{fig:fiber_mat}
  \end{minipage}
\end{figure}

Figure~\ref{fig:fiber_led} shows the average number of detected photons as a function of the illumination position along the fiber. Figure~\ref{fig:fiber_mat} compares the detected light yield for various surface treatments at the far end of the fiber, including perfect and partial mirrors, diffusers (Lambertian reflectors), and absorbers.

While small differences are observed between SPIROS and GEANT4, likely due to the mesh-based geometry representation in SPIROS (discussed further in Section~\ref{sec_performance}), the overall trends are well reproduced. These results support the correct implementation of photon generation in source mode and the accurate treatment of various surface types.

\section{Performance Evaluation}\label{sec_performance}

The computational performance of SPIROS is benchmarked against GEANT4 using the three detector models introduced in Section~\ref{sec_validation}. All simulations were executed in single-threaded mode on an Intel Core i7-11375H processor.

For GEANT4, two geometry implementations were tested: one using standard solid classes (such as G4Box and G4Tubs) as in Section~\ref{sec_validation}, and the other using G4TessellatedSolid to import the same mesh-based geometry as used in SPIROS. The execution speeds for each configuration are summarized in Table~1.

Across the three validation models, SPIROS was 2.1--2.7 times faster than GEANT4 with standard solid classes, and 3.6--9.1 times faster than GEANT4 with mesh-based geometry.
This performance gain arises from the lightweight architecture of SPIROS, which is tailored specifically for optical photon transport, and from efficient geometric intersection algorithms used for photon tracking, including hierarchical acceleration techniques described in Section~\ref{subsec_geometry}.

The performance of SPIROS is closely tied to the complexity of the imported mesh geometry, which is determined by the mesh resolution at the time of STL export. For the results in Section~\ref{sec_validation}, STL files were generated using the default mesh settings of Autodesk Inventor. While planar surfaces such as rectangular solids can be represented exactly with a small number of mesh triangles (e.g., 12 for a box), curved surfaces require finer tessellation to achieve accurate representation. This is especially relevant for configurations such as the cylindrical WLS fiber in Section~\ref{subsec_validation_fiber}, where internal photon propagation is sensitive to surface accuracy.

\begin{figure}[htbp]
    \begin{center}
        \includegraphics[width=0.54\linewidth]{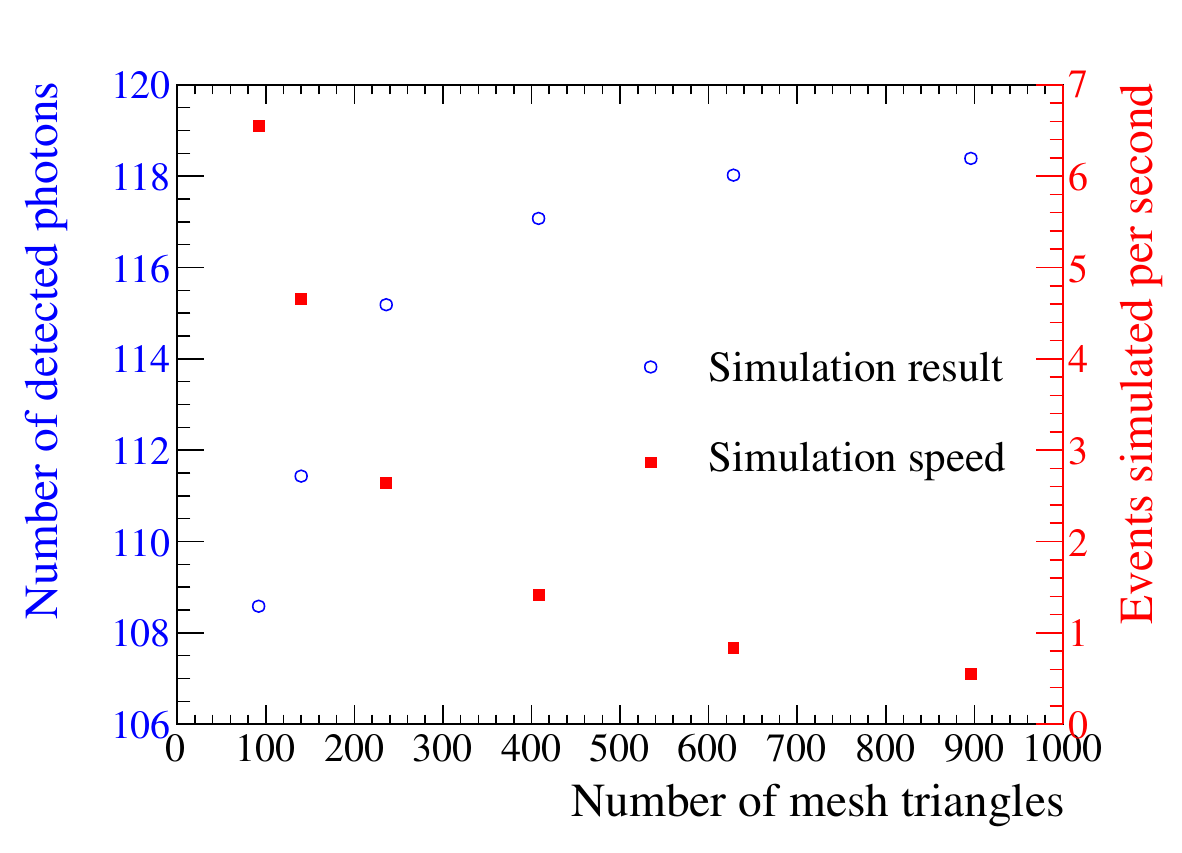}
         \caption{
            Simulation speed and result as a function of mesh granularity for the model of WLS fiber and LED described in Section~\ref{subsec_validation_fiber} when the LED-SiPM distance is 1.6~m. The default mesh setting contained 236 mesh triangles.
         }
         \label{fig:mesh}
    \end{center}
\end{figure}

As shown in Figure~\ref{fig:mesh}, as the number of mesh triangles increases, the simulation speed decreases inversely, while the light yield result converges and saturates beyond approximately 500 triangles.
This highlights the importance of appropriate mesh resolution to balance precision and computational efficiency.

For large-scale or high-statistics simulations, parallelization across multicore processors or computing clusters is effective. SPIROS supports this workflow by allowing users to launch independent jobs with different random seeds. Output files from separate runs can be seamlessly merged using the hadd utility provided by ROOT, enabling scalable performance without requiring specialized parallel programming.

\begin{table}[htbp]
  \centering
  \caption{Comparison of the average number of events simulated per second for the three validation models using SPIROS, GEANT4 with standard solid classes, and GEANT4 with mesh-based geometry. All simulations were performed in single-thread mode on an Intel Core i7-11375H processor.}
  \label{tab:time_comparison}
  \begin{tabular}{lrrr}
    \hline
    Detector model & SPIROS & Standard & Mesh-based \\
     &  & GEANT4 & GEANT4 \\
    \hline
    Scintillator + PMT (Sec.~\ref{subsec_validation_scinti}) & 48.8 & 23.5 & 13.5 \\
    Aerogel Cherenkov ring (Sec.~\ref{subsec_validation_cherenkov}) & 435 & 150 & 77.1 \\
    WLS fiber + LED (Sec.~\ref{subsec_validation_fiber}) & 2.64 & 0.99 & 0.29 \\
    \hline
  \end{tabular}
\end{table}

\section{Experimental Applications}\label{sec_application}
SPIROS has been utilized in the development, design optimization, and performance evaluation of several detectors in ongoing neutrino experiments such as T2K~\cite{bib:t2k}, NINJA~\cite{bib:ninja}, and AXEL~\cite{bib:axel}, as well as in exploratory research and development efforts for future neutrino experiments.
This section presents representative examples of how SPIROS has been applied in these contexts.

\subsection{Optical Response Study of the SuperFGD Detector in the T2K Experiment}
The SuperFGD detector~\cite{bib:tdr, bib:sfgd, bib:sfgd_kikawa, bib:sfgd_paper}, deployed in 2023 as part of the near detector upgrade in the T2K long-baseline neutrino oscillation experiment, consists of two million 1~cm$^3$ plastic scintillator cubes, each penetrated by three orthogonal WLS fibers that guide light to SiPMs. This unprecedented geometry posed unique challenges in understanding the optical behavior of the detector at early stages of development.

\begin{figure}[htbp]
    \begin{center}
        \includegraphics[width=1.\linewidth]{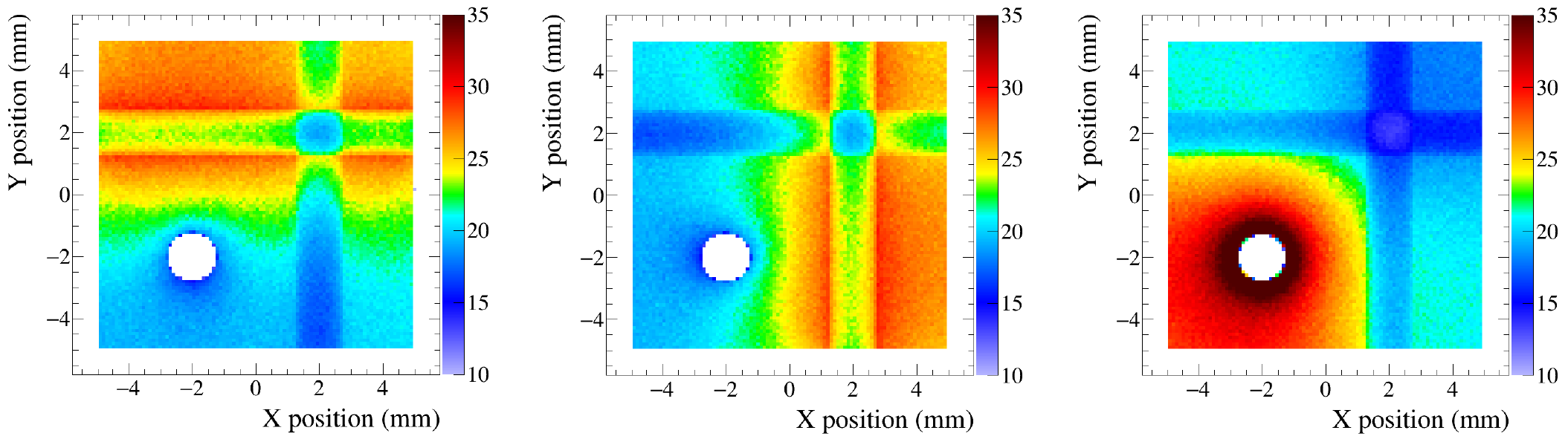}
         \caption{
            Simulated light yield from the scintillator cube in the X, Y, and Z fiber readouts (left, center, and right), shown as a function of the emission position.
         }
         \label{fig:cube}
    \end{center}
\end{figure}

SPIROS was used to simulate light propagation within individual cubes and light yield in each fiber readout. Figure~\ref{fig:cube} shows the simulated light yield in each fiber readout as a function of the emission position of a minimum ionizing particle (MIP). Although the light yield for each individual fiber exhibits strong spatial dependence, more than 15 photoelectrons are detected across all regions of the cube, ensuring sufficient signal for MIP detection and particle identification based on light yield. Moreover, the combined response from all three orthogonal fibers yields a consistently high and uniform light yield throughout the volume.

Subsequent detailed optical simulations using GEANT4, which accurately incorporated wavelength-dependent effects, and a positron beam test further refined the understanding of the detector’s response~\cite{bib:botao}. Nevertheless, the initial estimates obtained using SPIROS were found to be consistent with these more detailed studies.

\subsection{Optimization of Optical Interfaces in the SuperFGD Detector}
The optical interface between WLS fibers and SiPMs plays a crucial role in achieving high, stable, and uniform photon collection efficiency across the 55,888 readout channels of the SuperFGD detector.
The WLS fibers have a diameter of 1~mm, while the active area of each SiPM is 1.3~mm square.
Displacements between the fiber end and the SiPM surface can result in significant light loss.

\begin{figure}[htbp]
  \centering
  \begin{minipage}{0.47\textwidth}
    \centering
    \includegraphics[width=1.03\linewidth]{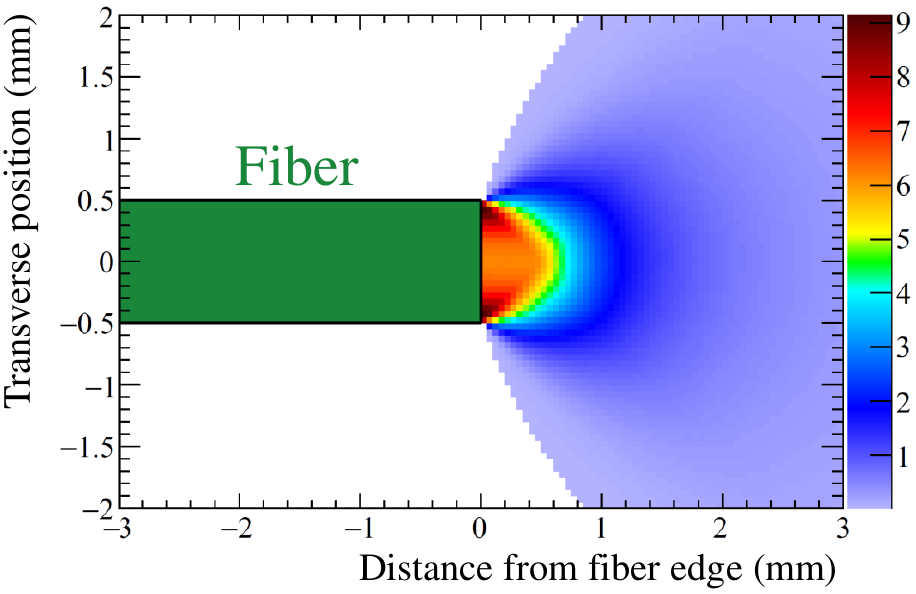}
    \caption{Optical photon distribution from the WLS fiber end.\newline}
    \label{fig:fiber_light}
  \end{minipage}
  \hfill
  \begin{minipage}{0.47\textwidth}
    \centering
    \includegraphics[width=0.95\linewidth]{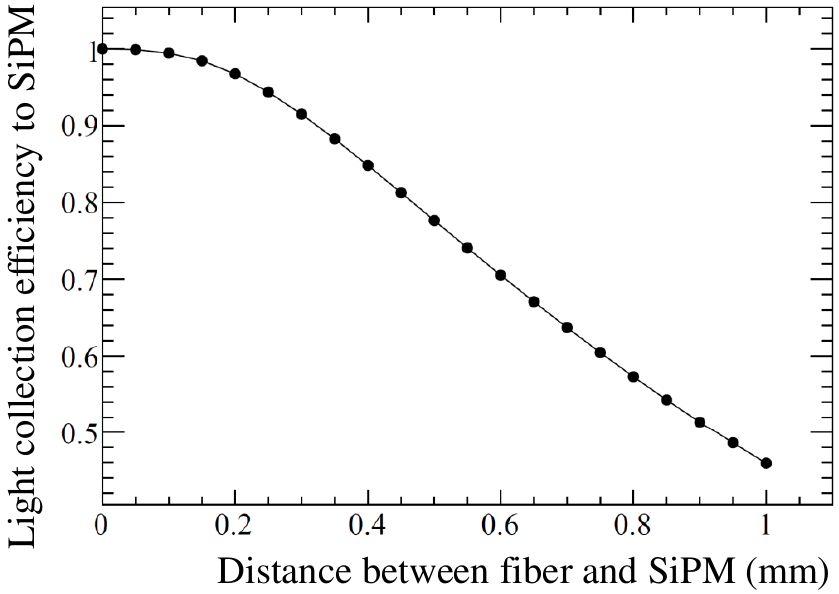}
    \caption{Light collection efficiency to the SiPM as a function of axial displacement between the fiber and the SiPM.}
    \label{fig:fiber_mppc}
  \end{minipage}
\end{figure}

SPIROS simulations were used to model the position and angular distribution of photons emerging from the fiber (Figure~\ref{fig:fiber_light}) and to evaluate the SiPM collection efficiency as a function of displacement between the fiber and the SiPM (Figure~\ref{fig:fiber_mppc}).
These results indicated that alignment accuracy within 200~$\mu$m is required to maintain sufficient light collection.
To address this requirement, a dedicated optical connector and a custom soft foam were designed and manufactured at CERN, with the connector ensuring precise alignment between the fibers and SiPMs, and the foam applying gentle pressure to eliminate any gap between them.

\subsection{Design Study of a Novel Scintillator Tracker for the NINJA Experiment}
The NINJA experiment aims to precisely measure neutrino interactions using nuclear emulsion detectors. To connect tracks in the emulsion detectors with a downstream muon range detector, a large-area, high-resolution scintillator tracker placed between the emulsion detectors and the muon range detector is required~\cite{bib:odagawa}.

\begin{figure}[htbp]
    \begin{center}
        \includegraphics[width=0.4\linewidth]{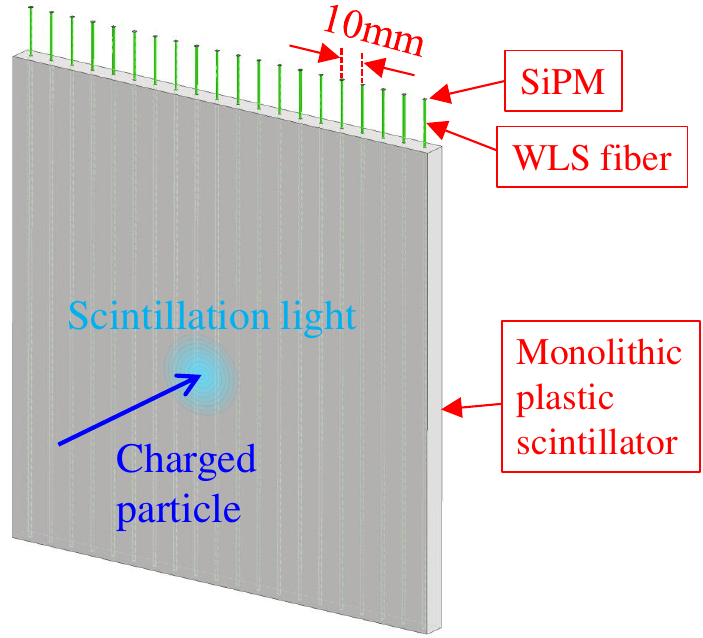}
         \caption{
            Conceptual illustration of the novel scintillator tracker consisting of a monolithic plastic scintillator slab with embedded WLS fibers arranged at regular intervals.
         }
         \label{fig:ninja_tracker}
    \end{center}
\end{figure}

I proposed a novel tracker concept consisting of a monolithic plastic scintillator slab with multiple embedded WLS fibers placed at regular intervals as shown in Figure~\ref{fig:ninja_tracker}. The position of charged particles can be inferred from the light yield distribution across the fibers, with higher signals expected from fibers closer to the track. SPIROS was utilized to optimize the detector design and evaluate the feasibility of this concept.

In this detector concept, achieving high light yield is essential to suppress statistical fluctuations in the detected photoelectron counts, and localizing light within the scintillator is also critical for accurate position reconstruction. First, a default configuration was simulated in SPIROS, with the characteristics of the scintillator, WLS fibers, and SiPMs based on those used in the T2K INGRID detector~\cite{bib:ingrid}. Various enhancements were then evaluated to estimate potential improvements in light yield. In addition, the impact of doping the scintillator with scattering agents to reduce the scattering length to 1~mm was studied to assess the effects of light localization.

For each configuration, reconstructed positions were obtained by fitting the simulated fiber signals with the particle position as a free parameter. The standard deviation of the residuals between true and reconstructed positions was used as a measure of positional resolution arising from statistical fluctuations.
The results are summarized in Table~\ref{tab:ninja}. Improvements in light yield led to better positional resolution. While increased scattering reduced the overall light yield, it significantly enhanced localization, and the positional resolution improved from 1.15~mm in the default configuration to 0.34~mm in the optimized case. Although this estimate reflects only statistical uncertainty under idealized conditions, the results motivated further studies toward realizing this detector concept, including light yield enhancement and the development of doped scintillator with scattering agents~\cite{bib:otani}.

\begin{table}[htbp]
  \centering
  \caption{Expected light yield and positional resolution in each configuration for the novel scintillator tracker.}
  \label{tab:ninja}
  \begin{tabular}{lrr}
    \hline
    Detector design & Total light & Positional \\
     & yield (P.E.) & resolution (mm) \\
    \hline
    (A) Default configuration &116 & 1.98\\
    (B) A + optical cement between scintillator and fiber &174 & 1.15\\
    (C) B + aluminum mirror deposition at fiber end &253 & 0.96\\
    (D) C + optical cement between fiber and SiPM &450 & 0.72\\
    (E) D + SiPM with higher photon detection efficiency &698 & 0.58\\
    (F) E + doping the scintillator with a scatterer &591 & 0.34\\
    \hline
  \end{tabular}
\end{table}

\subsection{Development of a Novel Water-Based Liquid Scintillator Detector for Future Neutrino Experiments}
\begin{figure}[htbp]
  \centering
  \begin{minipage}{0.47\textwidth}
    \centering
    \includegraphics[width=1.05\linewidth]{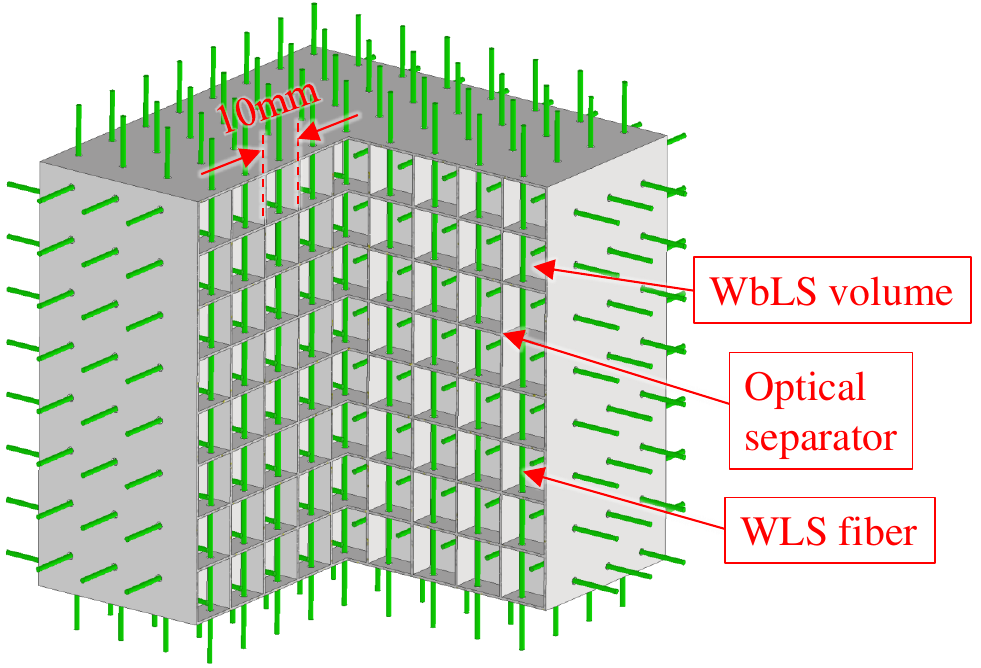}
    \caption{Conceptual illustration of the novel tracking detector based on WbLS, utilizing three-dimensional WLS-fiber readout.\newline}
    \label{fig:wbls_concept}
  \end{minipage}
  \hfill
  \begin{minipage}{0.47\textwidth}
    \centering
    \includegraphics[width=1.\linewidth]{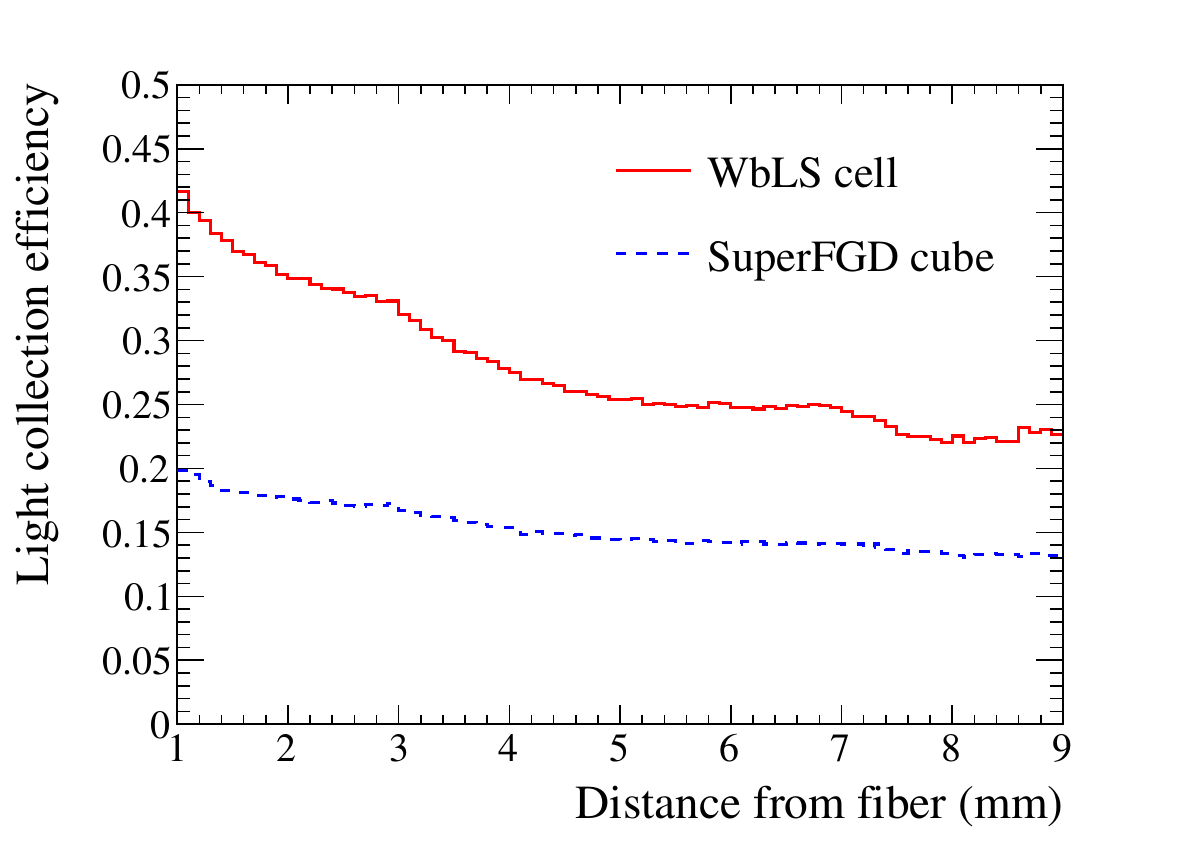}
    \caption{Comparison of light collection efficiency to a WLS fiber as a function of the distance from the scintillation point, for a WbLS cell and a SuperFGD cube.}
    \label{fig:wbls_result}
  \end{minipage}
\end{figure}

For future neutrino experiments using water Cherenkov detectors including Hyper-Kamiokande~\cite{bib:hk, bib:hk_es}, uncertainties in neutrino interactions on water target can introduce significant systematic errors. To mitigate this, I proposed a new detector concept based on water-based liquid scintillator (WbLS)~\cite{bib:yeh, bib:hans} segmented into 1 cm$^3$ cells using reflective materials. Each cell’s light is read out in three directions using WLS fibers as shown in Figure~\ref{fig:wbls_concept}.

A key concern was whether sufficient light yield could be achieved given the relatively low scintillation yield of WbLS compared to plastic scintillators or pure liquid scintillator. SPIROS simulations were used to estimate light collection efficiencies for this geometry. As shown in Figure~\ref{fig:wbls_result}, the simulated efficiency significantly exceeded that of the SuperFGD plastic scintillator design.

Further analysis revealed that this improvement arises from two factors: the absence of an air gap between the fiber and the scintillator, which reduces reflection losses, and the longer attenuation length of WbLS compared to plastic scintillators. These results demonstrated that the lower intrinsic scintillation yield of WbLS can be effectively mitigated by the enhanced collection efficiency, thereby supporting the potential of this detector concept and motivating its further development~\cite{bib:onda, bib:botao}.

\section{Conclusion}\label{sec_conclusion}

SPIROS is a lightweight and efficient optical simulation framework tailored for particle physics applications. It offers intuitive configuration, direct import of CAD-based detector geometries, and fast and accurate modeling of optical photon transport and interactions. Photon trajectories are tracked with full polarization and timing information, enabling detailed studies of detector response. The simulator supports both scintillation and Cherenkov processes, as well as photon generation from arbitrary sources.

Validation against GEANT4 was performed using three representative detector configurations: a scintillator bar with a PMT, a Cherenkov-ring imaging system with aerogel, and a WLS fiber illuminated by an LED. In all cases, SPIROS successfully reproduced the photon production, transport, and detection behavior with results in excellent agreement with those from GEANT4. These comparisons confirm that SPIROS reliably models key optical processes across diverse experimental setups.

Benchmark tests revealed that SPIROS significantly outperforms GEANT4 in terms of computational speed. Depending on the geometry, SPIROS achieved 2.1–2.7 times faster execution than GEANT4 using standard solids, and 3.6–9.1 times faster than GEANT4 with mesh-based geometry. This performance gain results from its lightweight architecture and optimized mesh intersection algorithms. The trade-off between simulation speed and mesh resolution was also investigated, showing that adequate precision can be achieved with a moderate number of mesh triangles, thereby offering a practical balance between accuracy and efficiency.

Finally, the applicability of SPIROS has been demonstrated through its use in various real-world detector studies, including the T2K SuperFGD, NINJA tracker, and a novel WbLS-based tracking detector. Its combination of physical fidelity, usability, and performance makes SPIROS a powerful tool for detector design, optimization, and prototyping in current and future particle physics experiments.

\section*{Acknowledgments}

I am deeply grateful to Tsuyoshi Nakaya for his continuous support, and insightful advice throughout this research.
I would like to express my sincere gratitude to Shinichi Akiyama, Soki Urano and Masashi Yoshida for using my software and providing valuable bug reports, and to Atsuko K. Ichikawa for introducing my software to them.



\begin{thebibliography}{99}
  \bibitem{bib_g4_1} S.~Agostinelli~\etal, Nucl. Instrum. Meth. A {\bf 506} 250-303 (2003)
  \bibitem{bib_g4_2} J.~Allison~\etal, IEEE Trans. Nucl. Sci. {\bf 53} 270-278 (2006)
  \bibitem{bib_g4_3} J.~Allison~\etal, Nucl. Instrum. Meth. A {\bf 835} 186-225 (2016)
  \bibitem{bib:assimp}ASSIMP: Open Asset Import Library. Available at: https://www.assimp.org
  \bibitem{bib:root_nim}R. Brun and F. Rademakers, Nucl. Instrum. Meth. A {\bf 389} 81–86 (1997)
  \bibitem{bib:root}ROOT: Data Analysis Framework, CERN. Available at: https://root.cern
  \bibitem{bib_mt}M.~Matsumoto and T.~Nishimura, ACM Trans. Model. Comput. Simul. {\bf 8}, 1, 3-30 (1998)
  \bibitem{bib_lambert} J. H.~Lambert. Photometria sive de mensure de gratibus luminis, colorum umbrae. Eberhard Klett (1760)
  \bibitem{bib:knoll}G. F.~Knoll, Radiation Detection and Measurement, Wiley, 4th ed. (2010)
  \bibitem{bib:frank_tamm}I. M.~Frank and I. Tamm, C.R. Acad. Sci. U.S.S.R., 14, 109 (1937)
  
  \bibitem{bib:bvh}S.~Gottschalk, M. C.~Lin, and D.~Manocha, Proc. ACM Siggraph'96, 171-180 (1996)

  \bibitem{bib:t2k} K.~Abe~\etal (T2K Collaboration), Nucl. Instrum. Meth. A {\bf 659} 106 (2011)
  \bibitem{bib:ninja}T.~Fukuda~\etal, Prog. Theor. Exp. Phys. {\bf 2017}, 6, 063C02 (2017)
  \bibitem{bib:axel}S.~Ban~\etal, Nucl. Instrum. Meth. A {\bf 875}, 185-192 (2017)
  
  \bibitem{bib:tdr} K.~Abe~\etal, arXiv:1901.03750 (2019)
  \bibitem{bib:sfgd} A.~Blondel~\etal, JINST {\bf 13} P02006 (2018)
  \bibitem{bib:sfgd_kikawa} T.~Kikawa~\etal, Nucl. Instrum. Meth. A {\bf 1080}, 170616 (2025)
  \bibitem{bib:sfgd_paper} S.~Abe~\etal, arXiv:2603.14921 (2026)
  \bibitem{bib:botao} S.~Abe~\etal, Nucl. Instrum. Meth. A {\bf 1080} 170757 (2025)
  \bibitem{bib:odagawa}T.~Odagawa~\etal, Nucl. Instrum. Meth. A {\bf 1034}, 166775 (2022)
\bibitem{bib:ingrid}K.~Abe~\etal (T2K Collaboration), Nucl. Instrum. Meth. A {\bf 694} 211-223 (2012)
  \bibitem{bib:otani}N.~Otani, arXiv:2501.04119 (2025)
  \bibitem{bib:hk} K.~Abe~\etal (Hyper-Kamiokande Proto-Collaboration), arXiv:1805.04163 (2018)
  \bibitem{bib:hk_es} K.~Abe~\etal (Hyper-Kamiokande Collaboration), arXiv:2506.16641 (2025)
  \bibitem{bib:yeh} M.~Yeh~\etal, Nucl. Instrum. Meth. A {\bf660}, 51 (2011)
  \bibitem{bib:hans} H. Th. J.~Steiger~\etal, arXiv:2405.05743 (2024)
  \bibitem{bib:onda}N.~Onta~\etal, Prog. Theor. Exp. Phys. {\bf 2025}, 12, 123H02 (2025)
  \bibitem{bib:botao}B.~Li~\etal, JINST {\bf21}, P01012 (2026)
\end{thebibliography}
\end{document}